# Quantum Plain and Carry Look-Ahead Adders

Kai-Wen Cheng  
u8911804@cc.nkfust.edu.tw

Chien-Cheng Tseng  
tcc@ccms.nkfust.edu.tw

Department of Computer and Communication Engineering,
National Kaohsiung First University of Science and Technology,
Yenchao, Kaohsiung, Taiwan, Republic of China

May 22, 2002

## Abstract

In this paper, two quantum networks for the addition operation are presented. One is the Modified Quantum Plain(MQP)adder, and the other is the Quantum Carry Look-Ahead(QCLA)adder. The MQP adder is obtained by modifying the Conventional Quantum Plain(CQP)adder. The QCLA adder is an extension of conventional digital Carry Look-Ahead adder. Compared with the CQP adder, two main advantages are as follows: First, the proposed MQP and QCLA adders have less number of elementary gates than the CQP adder. Secondly, the number of processing stages of the MQP and QCLA adder are less than ones of the CQP adder. As a result, the throughput time for computing the sum of two numbers on the quantum computer can be improved.

Keywords:Carry Look-Ahead Adder, Quantum Plain Adder, Quantum Arithmetic, parallel architecture

# 1. Introduction

In 1994, Shor showed that prime factorization and discrete logarithms can be done quickly on a quantum computer [1][2]. One application of Shor's quantum prime factorization algorithm is to break RSA cryptosystems. To implement Shor's algorithm, a lot of researchers have devoted themselves to design quantum arithmetic networks. In 1995, Vedral, Barenco and Ekert provided the quantum network for the modular exponentiation, which is one of the main arithmetic tasks of Shor's prime factorization algorithm [3]. To compute modular exponentiation, quantum networks of the plain adder, the adder modulo, and the multiplier modulo also proposed in [3]. Recently, Draper also proposed a method for addition on the quantum computer using the Quantum Fourier Transform [4]. Thus, to design an efficient quantum network to perform addition is an important research topic for implementing Shor's algorithm.

In this paper, two quantum adder networks are presented to improve the throughput time for computing the sum of two numbers on the quantum computer. One is the Modified Quantum Plain （MQP）adder, and the other is the Quantum Carry Look-Ahead（QCLA）adder. The paper is organized as follows: In section 2, the problem statement is made. In section 3, the conventional plain adder (CQP) in [3] is briefly reviewed and the MQP adder will be proposed. The number of the elementary gates and the number of processing stages are used to evaluate the performance of CQP and MQP adders. In section 4, the QCLA adder is designed and a comparison is made. The QCLA adder is an extension of conventional digital Carry Look-Ahead adder [5][6]. Finally, a conclusion is made.

# 2. Problem Statement

The addition is one of the most important arithmetic operations, so it is useful to design a quantum network to compute the sum of two numbers. Given two *n*-bit numbers ｜$a$〉 and ｜$b$〉 are as follows:

$$|a\rangle = |a_n a_{n-1} \ldots a_2 a_1\rangle \qquad (1)$$

$$|b\rangle = |b_n b_{n-1} \ldots b_2 b_1\rangle \qquad (2)$$

where ｜$a_i$〉 is the *i*th qubit of ｜$a$〉 and ｜$b_i$〉 is the *i*th qubit of ｜$b$〉. The purpose of this paper is to design a quantum network to compute the sum of ｜$a$〉 and ｜$b$〉. This sum has *n+1* qubits and is denoted by

$$|a+b\rangle = |C_n S_n S_{n-1} \ldots S_2 S_1\rangle \qquad (3)$$

where the qubit ｜$C_n$〉 represents the carry bit, and the qubit ｜$S_i$〉 is the sum bit. Let notation $\oplus$ be the Exclusive-OR operation, then the carry bit ｜$C_n$〉 can be computed by the following recursive relation [5][6]:

$$C_i = a_i b_i \oplus (a_i \oplus b_i) C_{i-1} \qquad \text{and} \qquad C_0 = 0 \qquad (4)$$

and the relationship among ｜$a_i$〉, ｜$b_i$〉 and ｜$S_i$〉 is given by

$$S_i = (a_i \oplus b_i) \oplus C_{i-1} \qquad (5)$$

The quantum system block diagram to perform the addition is shown in Figure 1. The input contains three components as $|a\rangle$, $|b\rangle$, and one $(n+1)$-bit zero number $|0\rangle$. The output of network includes one number $|a\rangle$, the sum $|S\rangle = |S_n S_{n-1}...S_2 S_1\rangle$ and $(n+1)$-bit carry $|C\rangle = |C_n C_{n-1}...C_2 C_1 C_0\rangle$. Clearly, the sum $|S\rangle$ appeared in the same qubits represented the number $|b\rangle$. Note that only the carry bit $|C_n\rangle$ is what to be considered, and the other qubits of $|C\rangle$ are the byproduct of the addition arithmetic. Now, the problem is how to design an efficient addition quantum network in Figure 1 such that the throughput time of addition operation can be improved. In next section, the CQP adder is first reviewed. Then, the number of elementary gates and processing stages are used to evaluate the performance of the CQP adder.

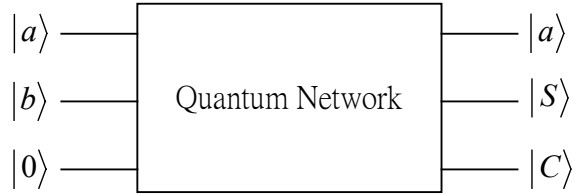

Figure 1 The quantum system block diagram of addition for two numbers a and b.

## 3. Quantum Plain Adder

### 3.1 Conventional Quantum Plain Adder

The Conventional Quantum Plain (CQP) adder [3] is implemented with two unitary computational gates. One is a carry gate in Figure 2, and the other is a sum gate in Figure 3. In this paper, the carry gate with a dark bar on the left side is used to represent the reverse order operation of a carry gate with a dark bar on the right side. From Figure 2, we see that the fourth output qubit of the carry gate is equal to

$$((a_i b_i) \oplus C_i) \oplus ((a_i \oplus b_i)C_{i-1}) \qquad (6)$$

If the initial input value of $C_i$ is set to zero, the output relation in (6) is reduced to the one in (4). Thus, the information of carry is placed in the fourth qubit of the output of the carry gate, and the information $a_i \oplus b_i$ is located in the third qubit of the output of the carry gate, while the first two qubits of the carry gate keep unchanged. In Figure 3, the sum gate implements the relation in (5). The information of sum appears in the third output qubit of the sum gate. However, the first two qubits of the sum gate are the same as its input.

The CQP adder composed of the previous unitary gates for two 4-bit numbers $|a\rangle$ and $|b\rangle$ is shown in Figure 4. The Figure 5 is obtained from Figure 4 by replacing the sum and carry gates with their details. The CQP adder used $3n+1$ qubits to add two n-bit numbers. The reason is that the CQP Adder needs additional $n$ temporary qubits to process carry information. Besides, the initial states of each temporary carry qubit in the input are set to zero, i.e. $|C_i\rangle = 0$, for $i = 0,1,2,3,4$. When the computation of the addition is completed, the carry qubits are reset to zero except $|C_n\rangle$. The most significant qubit of carry $|C_n\rangle$ is the last qubit of the quantum network in Figure 4. Furthermore, the input qubit $|b_i\rangle$ is replaced with sum bit $|S_i\rangle$, for $i = 1,2,3,4$. In next subsection, the performance of this CQP adder will be evaluated.

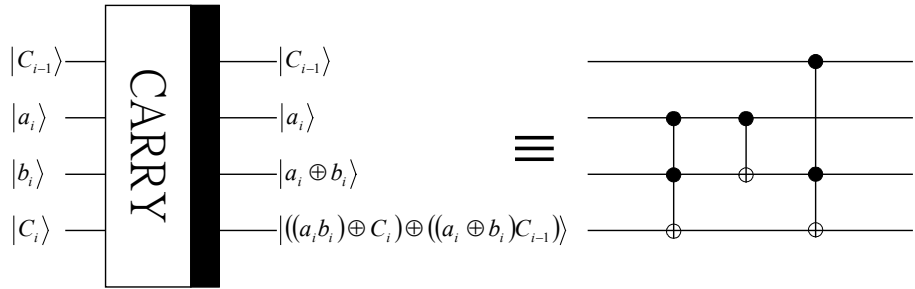

Figure 2  Carry gate.

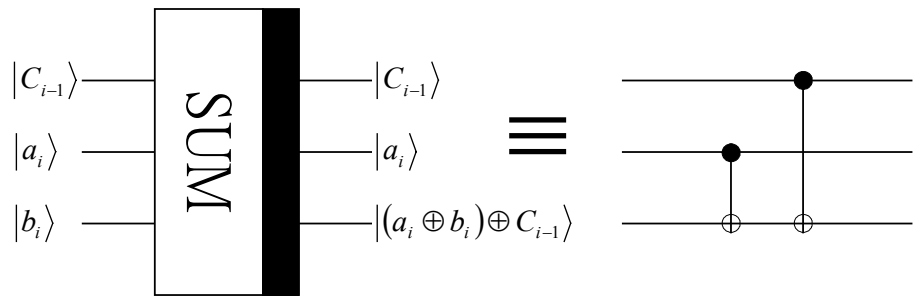

Figure 3  Sum gate.

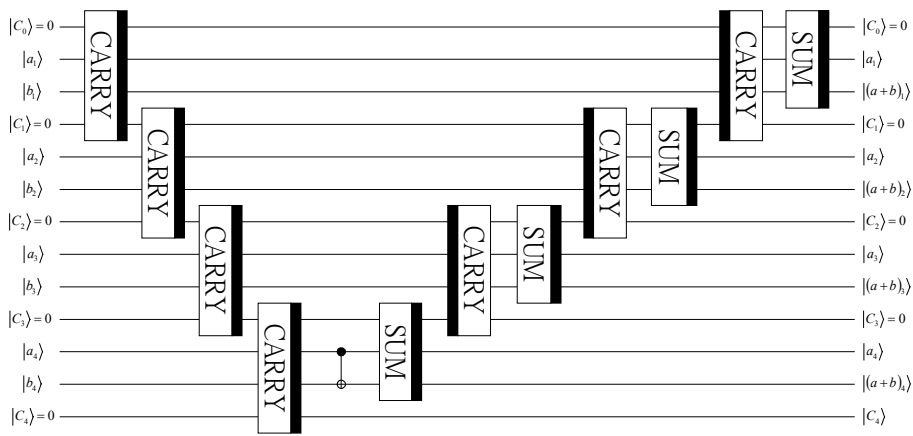

Figure 4  The Conventional Quantum Plain adder for two 4-bit numbers.

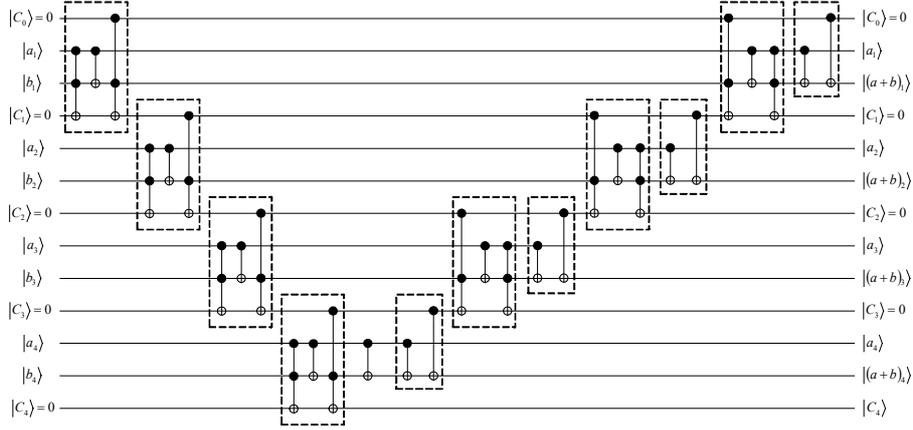

Figure 5 The Conventional Quantum Plain adder for two 4-bit numbers.

## 3.2 Performance Evaluation of the Conventional Quantum Plain Adder

In order to evaluate the performance of CQP adder, we consider the following two factors. One is the number of processing stages from input to output, and the other is the number of Controlled-Not (CNOT) gate and Controlled-Controlled-Not (CCNOT) gate. When the number of processing stages is small, it means that the throughput time is small. Besides, the smaller number of elementary gates is, the smaller complexity of adder has. For counting the minimum number of processing stages of CQP adder in Figure 5, the first two elementary gates CCNOT gate and CNOT gate of each Carry gates in Figure 5 are moved toward the input because independent gates can be grouped together to process such that the throughput time can be reduced. Therefore, the quantum network in Figure 5 can be rearranged into Figure 6. Both networks perform the same function. The only difference between them is the order of processing. From the help of auxiliary vertical dashed lines in Figure 6, it is clear that the number of processing stages of the CQP Adder for two 4-bit numbers is 24.

The elementary gates of CQP adder are the CNOT gate or the CCNOT gate used either within a carry gate and a sum gate or outside. Each elementary gate is counted with one gate. Figure 7 is exactly same as Figure 5 except auxiliary vertical dashed lines are inserted to help us to count the number of gates. From Figure 7, it is clear that the number of gates of the CQP adder for two 4-bit numbers is 30. Moreover, the CQP adder is designed to compute exponentiation modulo [3], so the temporary bits of carry need to be reset to zero. If we only want to perform addition operation, the action to reset carry bits can be removed such that the CQP adder can be simplified. In the following subsection, a CQP adder without resetting the temporary bits of carry will be presented.

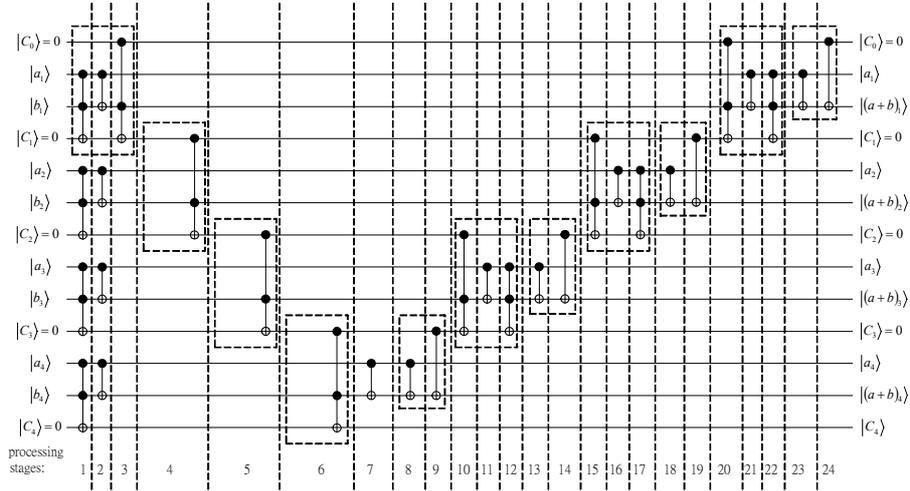

Figure 6  The number of processing stages of a CQP adder for two 4-bit numbers.

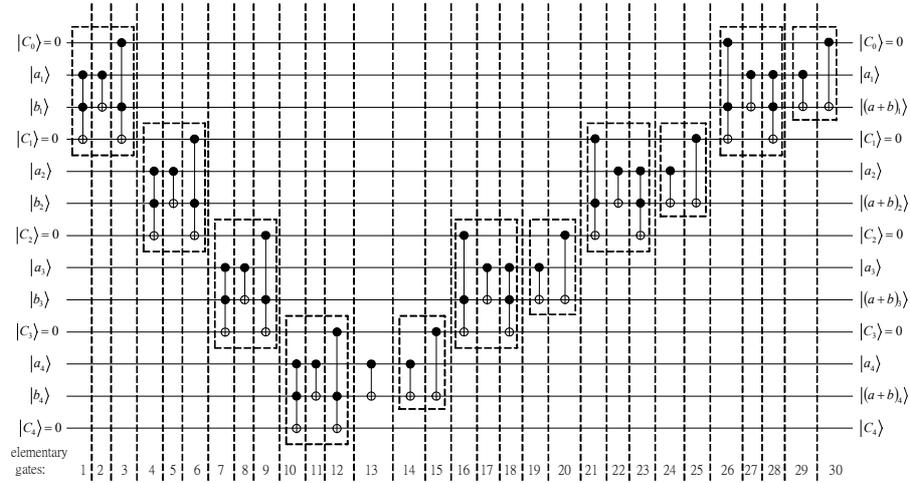

Figure 7  The number of elementary gates of a CQP adder for two 4-bit numbers.

### 3.3  Modified quantum plain adder without resetting temporary bits of carry to zero

The CQP adder in Figure 5 can be modified into the quantum adder network in Figure 8. The inverse Carry gates and the first CNOT gate of each Sum gate in Figure 4 and 5 are removed. The Modified Quantum Plain (MQP) adder does not need to reset temporary bits of carry to zero, so the number of elementary gates can be reduced. Now, the performance of the MQP adder is evaluated. For counting the minimum number of processing stages, the first two elementary gates CCNOT gate and CNOT gate of each Carry gates in Figure 8 are moved toward the input, and the four CNOT gates outside of the carry gates are gathered. Therefore, the quantum network in Figure 8 is rearranged into Figure 9. Clearly, the number of processing stages of the MQP Adder for two 4-bit numbers is 7. In addition, from the Figure 10, we see that the number of elementary gates of the MQP Adder for two four-bit numbers is 16.

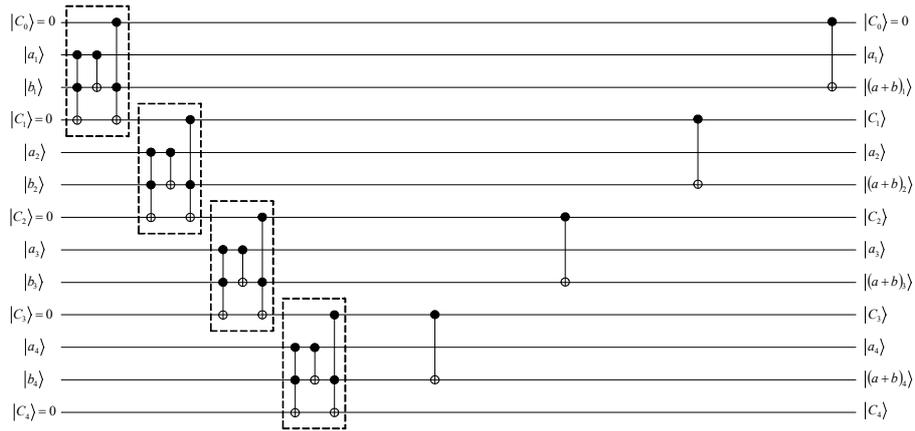

Figure 8  The Modified Quantum Plain adder for two 4-bit numbers.

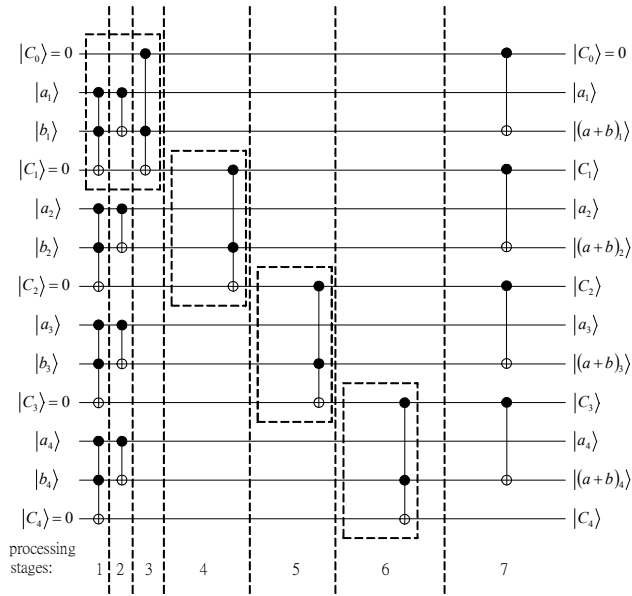

Figure 9  The number of processing stages of a MQP Adder for two 4-bit numbers.

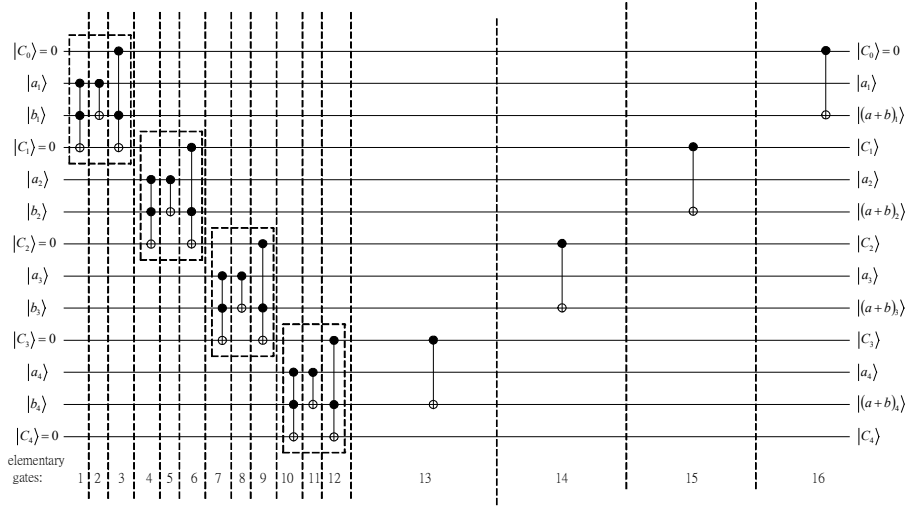

Figure 10　The number of elementary gates of a MQP Adder for two 4-bit numbers.

# 4. Quantum Carry Look-Ahead Adder

### 4.1　Quantum Carry Look-Ahead Adder

According to （4） and （5）, the recursive relations of carry bit and sum bit are rewritten as follows:

$$|C_i\rangle = |a_ib_i + (a_i \oplus b_i)C_{i-1}\rangle = |a_ib_i \oplus (a_i \oplus b_i)C_{i-1}\rangle = |g_i \oplus p_iC_{i-1}\rangle \quad (7)$$

$$|S_i\rangle = |(a_i \oplus b_i) \oplus C_{i-1}\rangle = |p_i \oplus C_{i-1}\rangle \quad (8)$$

where $g_i = a_ib_i$ is used to generate the carry bit $|C_i\rangle$, and $p_i = a_i \oplus b_i$ is used to propagate the previous carry bit $|C_{i-1}\rangle$ to the present carry bit $|C_i\rangle$ or sum bit $|S_i\rangle$. The Quantum Carry Look-Ahead (QCLA) adder is created by expanding （7） and （8） into the following form:

$$\begin{aligned}
|C_1\rangle &\to |g_1 \oplus p_1C_0\rangle = |C_1\rangle \\
|C_2\rangle &\to |g_2 \oplus p_2C_1\rangle = |g_2 \oplus g_1p_2 \oplus p_2p_1C_0\rangle = |C_2\rangle \\
|C_3\rangle &\to |g_3 \oplus p_3C_2\rangle = |g_3 \oplus g_2p_3 \oplus g_1p_2p_3 \oplus p_3p_2p_1C_0\rangle = |C_3\rangle \\
&\vdots \\
|C_n\rangle &\to |g_n \oplus p_nC_{n-1}\rangle \\
&= |g_n \oplus g_{n-1}p_n \oplus g_{n-2}p_{n-1}p_n \oplus g_{n-3}p_{n-2}p_{n-1}p_n \oplus \cdots \oplus p_np_{n-1}\cdots p_2p_1C_0\rangle = |C_n\rangle
\end{aligned} \quad (9)$$

$$|b_1\rangle \rightarrow |p_1 \oplus C_0\rangle = |S_1\rangle$$
$$|b_2\rangle \rightarrow |p_2 \oplus C_1\rangle = |p_2 \oplus g_1 \oplus p_1 C_0\rangle = |S_2\rangle$$
$$|b_3\rangle \rightarrow |p_3 \oplus C_2\rangle = |p_3 \oplus g_2 \oplus g_1 p_2 \oplus p_2 p_1 C_0\rangle = |S_3\rangle \quad (10)$$
$$|b_4\rangle \rightarrow |p_4 \oplus C_3\rangle = |p_4 \oplus g_3 \oplus g_2 p_3 \oplus g_1 p_2 p_3 \oplus p_3 p_2 p_1 C_0\rangle = |S_4\rangle$$
$$\vdots$$
$$|b_n\rangle \rightarrow |p_n \oplus C_{n-1}\rangle$$
$$= |p_n \oplus g_{n-1} \oplus g_{n-2} p_{n-1} \oplus g_{n-2} p_{n-2} p_{n-1} \oplus g_{n-4} p_{n-3} p_{n-2} p_{n-1} \oplus \cdots$$
$$\oplus p_{n-1} p_{n-2} \cdots p_2 p_1 C_0\rangle = |S_n\rangle$$

In (9) and (10), the symbol $\rightarrow$ represents the transform relationship of state from its left side (input) to its right side (output) on a specific qubit. In (9), the carry bit $|C_i\rangle$ is located at the same qubit represented its initial state $|0\rangle$ in input. However, in (10), the qubit $|b_i\rangle$ is replaced with the sum $|S_i\rangle$. In the following, the QCLA adder will be constructed based on (9) and (10).

First, let us define the AND gate in Figure 11, and the XOR gate in Figure 12. Moreover, based on (9), a $C_i$ module is constructed in Figure 13, and from (10), a $S_i$ module is built in Figure 14. The AND gate is composed of a CCNOT gate, and it is represented the AND operation of the *ith*-bit of two numbers $a$ and $b$, i.e. $|g_i\rangle = |a_i b_i\rangle$. In Figure 11, the third qubit of input is set to zero, so that the corresponding output is the AND operation of the first two qubits of input. The XOR gate consists of a CNOT gate, and it stands for the Exclusive-OR operation of the *ith*-bit of two numbers $a$ and $b$, i.e. $|p_i\rangle = |a_i \oplus b_i\rangle$. In Figure 12, the second qubit of output is the Exclusive-OR operation of two input numbers. The $C_i$ module in Figure 13 and the $S_i$ module in Figure 14 are composed of the CNOT gate, the CCNOT gate, and some multiple-Controlled NOT gate [7]. What we considered is the MSB of Carry $C_n$, and it is given by $C_i$ module with $i=n$. The $S_i$ module is the general form to get sum bit for $i=1,2,...,n$. Using basic gates and modules from Figure 11 to Figure 14, the QCLA adder for two *n*-bit numbers is depicted in Figure 15. Figure 15(b) is obtained from Figure 15(a) by replacing AND gate, XOR gate, $C_i$ module and $S_i$ module by their details. The QCLA adder is designed only for the purpose of addition, so that it is not required to reset the temporary bits of carry to zero. The QCLA adder is a parallel architecture by sharing the outputs of the AND gate and the XOR gate in Figure 15(a). In the following subsection, the performance of the QCLA adder is evaluated.

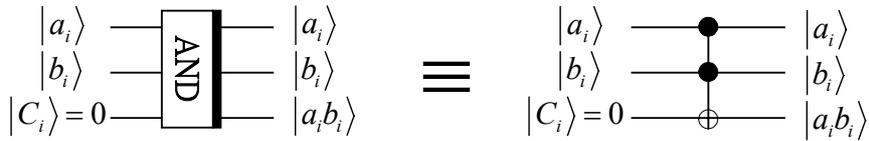

Figure 11  AND gate.

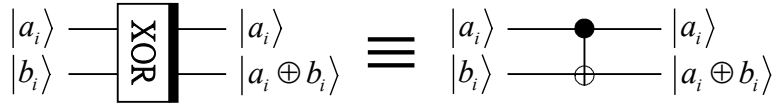

Figure 12  XOR gate.

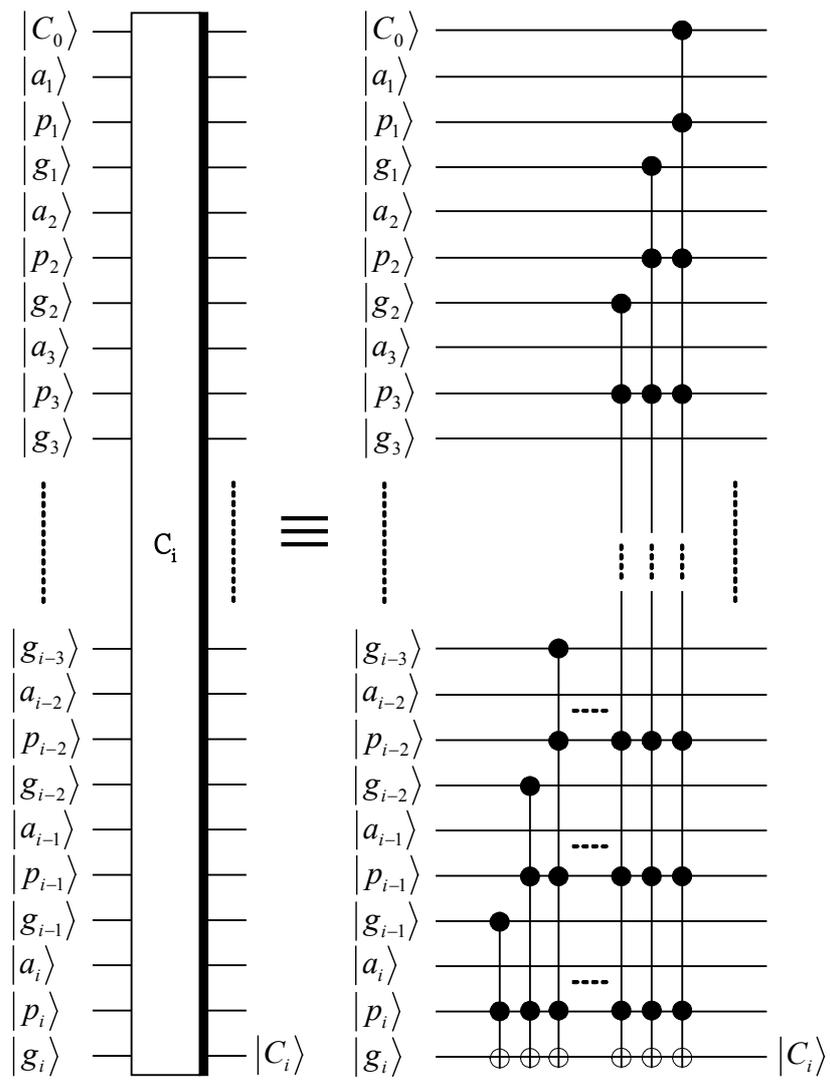

Figure 13　$C_i$ module of the QCLA adder.

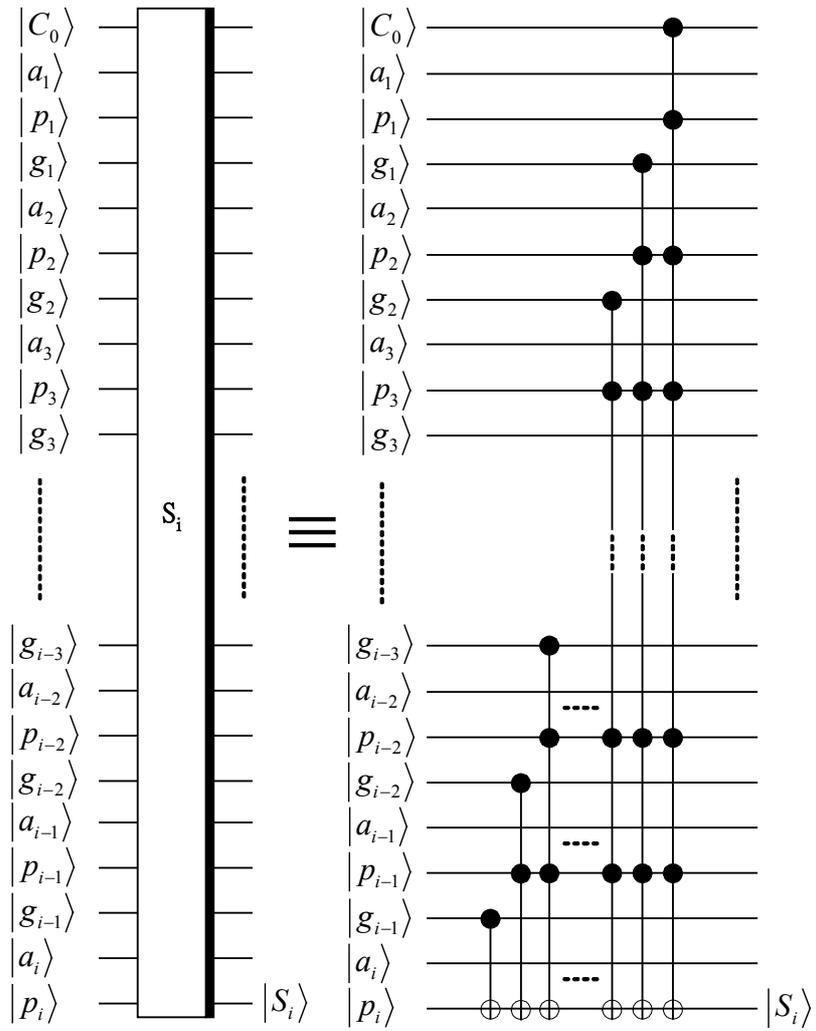

Figure 14  $S_i$ module of the QCLA adder.

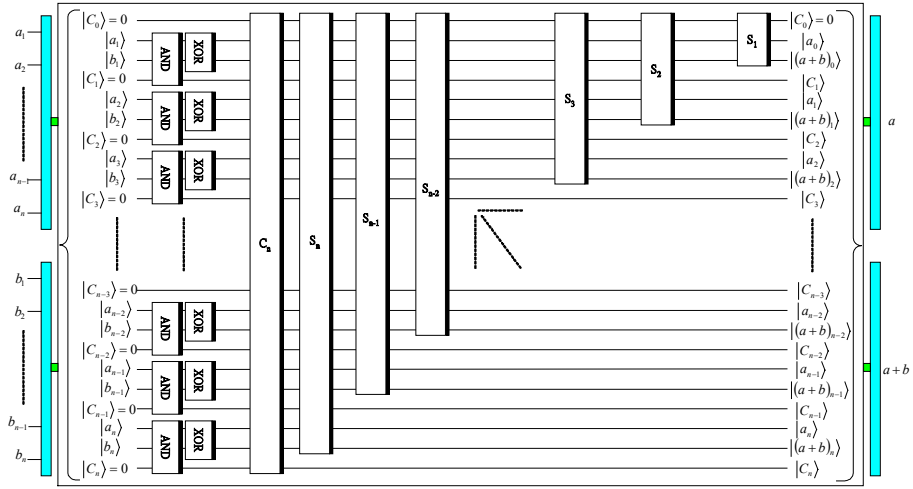

Figure 15(a) The QCLA adder for two n-bit numbers without resetting temporary bits of carry.

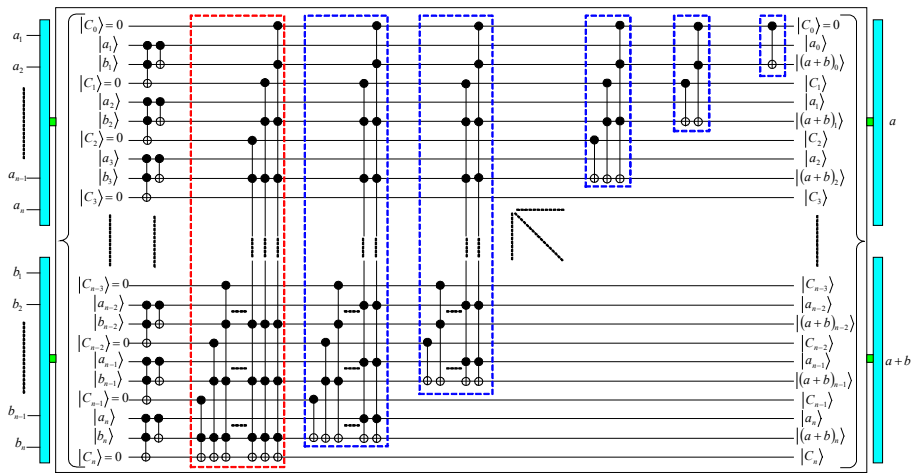

Figure 15(b) The QCLA adder for two n-bit numbers without resetting temporary bits of carry.

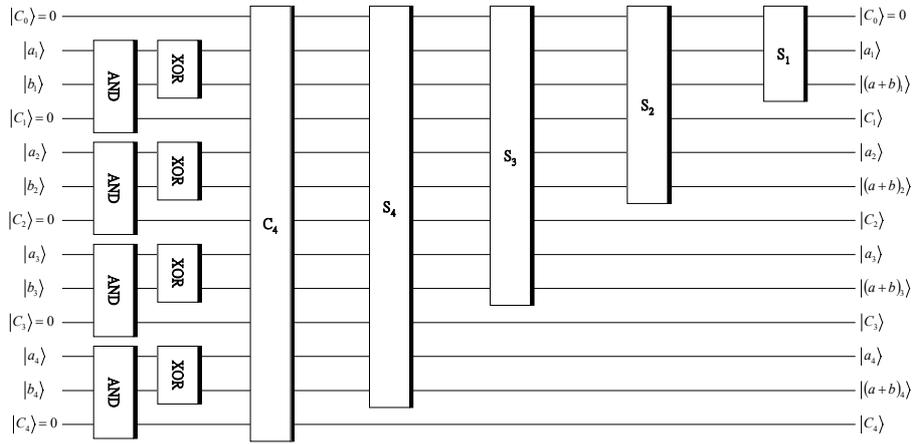

Figure 16(a)   The QCLA adder for two 4-bit numbers without resetting temporary bits of carry.

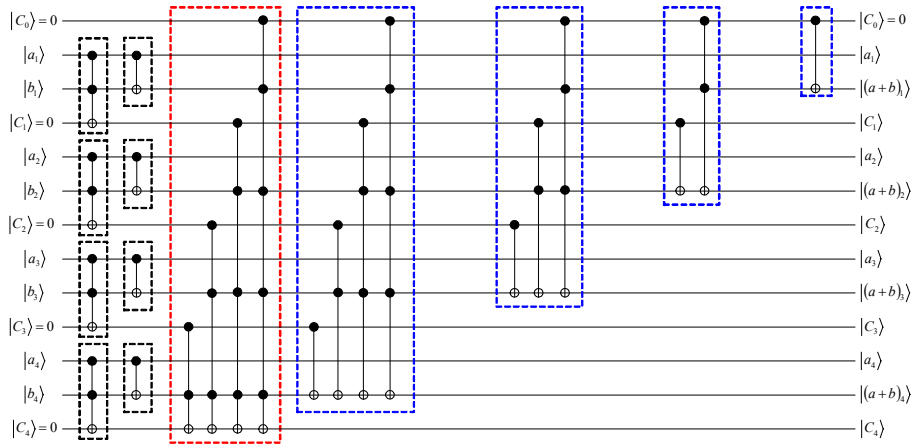

Figure 16(b)   The QCLA adder for two 4-bit numbers without resetting temporary bits of carry.

### 4.2 Performance Evaluation of the Quantum Carry Look-Ahead Adder

For convenience of performance evaluation, Figure 15 with *n=4* is shown in Figure 16. The elementary gates of CQP adder are the CNOT gate and the CCNOT gate, however, the elementary gates of QCLA adder includes the *n*-times Controlled-NOT gate, i.e. the $C^iNOT$ gate for *i=1,2,...,n*. The number of processing stages of the QCLA adder for two four-bit numbers is 6, as shown in figure 17. From left side to right side in Figure 17, the first gate we encountered is the AND gate. Here, we have four AND gates. These four AND gate are counted as one processing stage, because these gates can be processed at the same time. For the same reason, the four XOR gates are counted as one processing stage. Then, we consider the processing stages within the $C_i$ module for *i =4* because we only care the longest throughput time for the addition operation. Therefore, the $C_4$ module is what we consider. In Figure 17, the $C_4$ module consists of a CCNOT gate, a $C^3$NOT gate, a $C^4$NOT gate, and a $C^5$NOT gate. Each gate in the $C_4$ module is counted as one processing stage. So, the total processing stages is 6.

From Figure 18, we see that the number of elementary gates of the QCLA adder for two four-bit numbers is 22. The elementary gates in OCLA adder includes the $C^iNOT$ gate for *i=1,2,...,n* used either within AND gate, XOR gate, $C_i$ module, and $S_i$ module or outside. Finally, for *n-bit* addition, it can be shown that the number of processing stages for a QCLA adder is n＋2, and the number of elementary gates for it is $4n+\Sigma_{i=1\sim n-1}(n-i)$. Moreover, the number of processing stages for a CQP adder is 6n, and the number of elementary gates for a CQP adder of is 8n－2. Besides, the number of processing stages for a MQP adder is n＋3, and the number of elementary gates for it is 4n. The comparison among QCLA adder, CQP adder, and MQP adder is summarized in Table 1.

From Table 1, it is clear that the number of processing stages for a QCLA adder is less than the CQP adder and MQP adder, i.e. the throughput time of a QCLA adder is improved compared with CQP adder and MQP adder. However, the number of elementary gates of a QCLA adder is less than the CQP adder but more than the MQP adder.

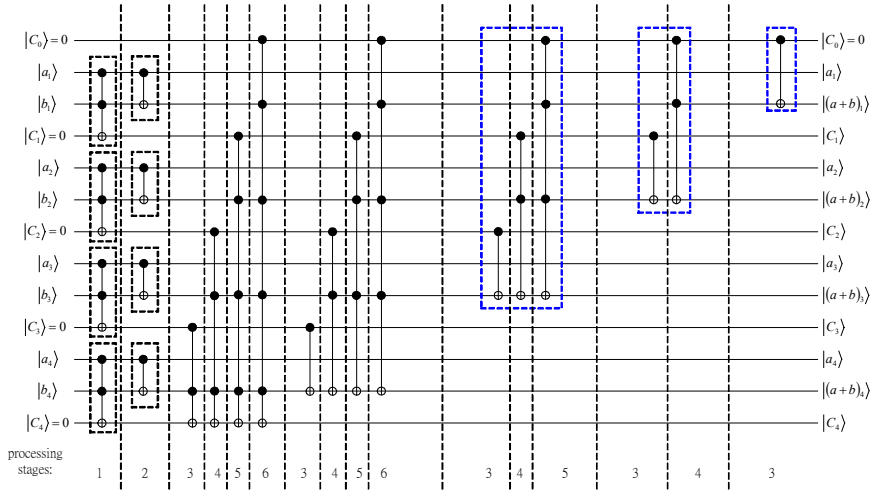

Figure 17 The number of processing stages of a QCLA adder for two 4-bit numbers without resetting temporary bits of carry.

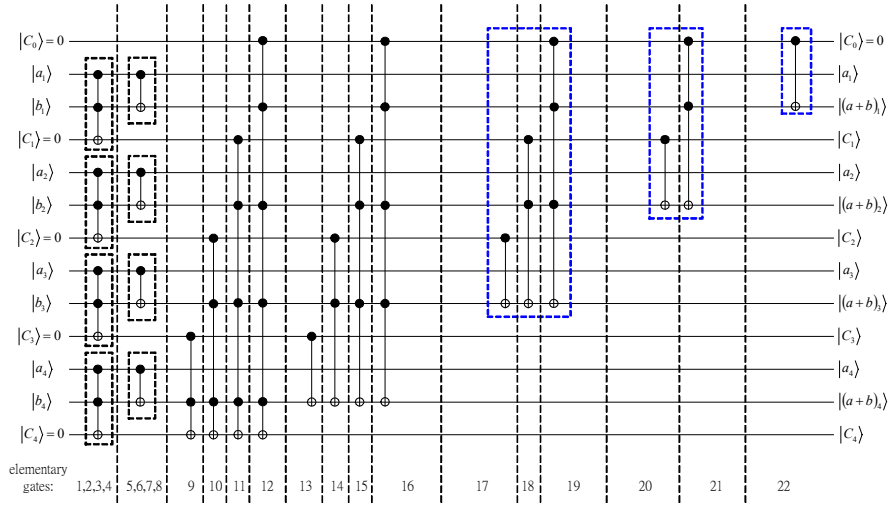

Figure 18 The number of elementary gates of a QCLA adder for two 4-bit numbers without resetting temporary bits of carry.

# 5. Conclusion

In this paper, two quantum networks for the addition operation have been presented. One is the MQP adder, and the other is the QCLA adder. Compared with the CQP adder, two main advantages are as follows: First, the proposed MQP and QCLA adders have less number of elementary gates than the CQP adder. Secondly, the number of processing stages of the MQP and QCLA adder are less than ones of the CQP adder. As a result, the throughput time for computing the sum of two numbers on the quantum computer can be improved.

With the throughput time of the addition on a quantum computer is improved, the arithmetic quantum networks can be realized faster. One of the applications is that we can design a faster quantum network system to factoring a composite number, i.e. breaking the RSA cryptosystems with Shor's Quantum Factoring Algorithm on a quantum computer [8][9][10] can be speeded up.

| Number of n | QCLA Adder | | CQP Adder | | MQP Adder | |
|---|---|---|---|---|---|---|
| | Number of Processing Stages | Number of Elementary Gates | Number of Processing Stages | Number of Elementary Gates | Number of Processing Stages | Number of Elementary Gates |
| 1 | 3 | 4 | 6 | 6 | 4 | 4 |
| 2 | 4 | 9 | 12 | 14 | 5 | 8 |
| 3 | 5 | 15 | 18 | 22 | 6 | 12 |
| 4 | 6 | 22 | 24 | 30 | 7 | 16 |
| ⋮ | ⋮ | ⋮ | ⋮ | ⋮ | ⋮ | ⋮ |
| n | n+2 | $4n+\Sigma_{i=1\sim n-1}(n-i)$ | 6n | 8n−2 | n+3 | 4n |

Table 1　The Comparison among the QCLA Adder, the CQP Adder, and the MQP Adder.